\documentclass{PHEA}  
\usepackage{graphicx}
\usepackage{txfonts}
\usepackage{lipsum}
\usepackage{subcaption}         % necessary for continued figures, example in section 3
\usepackage{lscape}             % to rotate a single page table, example in appendix.
\usepackage{placeins}           % useful with \FloatBarrier, to keep 
\def\apjl{ApJ}                % Astrophysical Journal, Letters
\def\apjs{ApJS}               % Astrophysical Journal, Supplement
\def\ao{Appl.~Opt.}           % Applied Optics
\def\baas{\ref@jnl{BAAS}}               % Bulletin of the AAS
\def\bac{\ref@jnl{Bull. astr. Inst. Czechosl.}}
\def\caa{\ref@jnl{Chinese Astron. Astrophys.}}
\def\cjaa{\ref@jnl{Chinese J. Astron. Astrophys.}}
\def\icarus{\ref@jnl{Icarus}}           % Icarus
\def\jcap{\ref@jnl{J. Cosmology Astropart. Phys.}}
\def\jrasc{\ref@jnl{JRASC}}             % Journal of the RAS of Canada
\def\memras{\ref@jnl{MmRAS}}            % Memoirs of the RAS
\def\mnras{\ref@jnl{MNRAS}}             % Monthly Notices of the RAS
\def\na{\ref@jnl{New A}}                % New Astronomy
\def\nar{\ref@jnl{New A Rev.}}          % New Astronomy Review
\def\pra{\ref@jnl{Phys.~Rev.~A}}        % Physical Review A: General Physics
\def\prb{\ref@jnl{Phys.~Rev.~B}}        % Physical Review B: Solid State
\def\prc{\ref@jnl{Phys.~Rev.~C}}        % Physical Review C
\def\prd{\ref@jnl{Phys.~Rev.~D}}        % Physical Review D
\def\pre{\ref@jnl{Phys.~Rev.~E}}        % Physical Review E
\def\prl{\ref@jnl{Phys.~Rev.~Lett.}}    % Physical Review Letters
\def\pasa{\ref@jnl{PASA}}               % Publications of the Astron. Soc. of Australia
\def\pasp{\ref@jnl{PASP}}               % Publications of the ASP
\def\pasj{\ref@jnl{PASJ}}               % Publications of the ASJ
\def\rmxaa{\ref@jnl{Rev. Mexicana Astron. Astrofis.}}%
\def\qjras{\ref@jnl{QJRAS}}             % Quarterly Journal of the RAS
\def\zap{\ref@jnl{ZAp}}                 % Zeitschrift fuer Astrophysik
\def\jqsrt{\ref@jnl{J.~Quant.~Spec.~Radiat.~Transf.}}
\def\memsai{\ref@jnl{Mem.~Soc.~Astron.~Italiana}}
\def\nphysa{\ref@jnl{Nucl.~Phys.~A}}   % Nuclear Physics A
\def\physrep{\ref@jnl{Phys.~Rep.}}   % Physics Reports
\def\physscr{\ref@jnl{Phys.~Scr}}   % Physica Scripta
\def\planss{\ref@jnl{Planet.~Space~Sci.}}   % Planetary Space Science
\def\procspie{\ref@jnl{Proc.~SPIE}}   % Proceedings of the SPIE

\usepackage{xcolor}

\usepackage{array,tabularx,booktabs,colortbl}

\newcolumntype{Y}{%
  >{\raggedright\arraybackslash}X}

\definecolor{tableheader}{RGB}{225,236,246}
\definecolor{tablealt}{RGB}{247,250,253}
\newcommand{\re}[1]{#1}

\begin{document}

\pheaNum{XXXXXX}
%%%%%%%%%%%%%%%%%%%%%%%%%%%%%%%%%%%%%%%%
% if you use custom commands in your title,
% ensure to check your title when submitting!
%%%%%%%%%%%%%%%%%%%%%%%%%%%%%%%%%%%%%%%%
   \title{The Galactic Cosmic-Ray Knee in the LHAASO Era}
%%%%%%%%%%%%%%%%%%%%%%%%%%%%%%%%%%%%%%%%
% Please separate each author with the \and command
%
% Use the \corrauth to provide the corresponding
% author address. It will be automatically inserted as 
% footnote in the PDF output.
%
% Please DO NOT include ORCIDs next to author names.
% Instead, please provide an active address for each coauthor:
% it will be automatically extracted by EDPS editorial system, 
% and co-authors will be be able to authenticate their ORCID.
%
% Only authenticated ORCIDs will be taken into account.
% ORCIDs included here will be removed.
%%%%%%%%%%%%%%%%%%%%%%%%%%%%%%%%%%%%%%%%

   \author{Bing Theodore Zhang\inst{1,2}\corrauth{zhangbing@ihep.ac.cn}        % use \corrauth for the corresponding author
        \and Zhen Cao\inst{1,2,  3}\corrauth{caozh@ihep.ac.cn} 
        }

  \institute{Key Laboratory of Particle Astrophysics and Experimental Physics Division and Computing Center, Institute of High Energy Physics, Chinese Academy of Sciences, 100049 Beijing, China
\and 
TIANFU Cosmic Ray Research Center, 610213,Chengdu, Sichuan , China
\and
University of Chinese Academy of Sciences, 100049 Beijing, China
}

% \abstract{}{}{}{}{}
% 5 {} token are mandatory
 
  \abstract
   {The origin of the Galactic cosmic-ray (CR) knee remains unresolved, largely because measurements of individual nuclear spectra around the knee have been limited and direct observational constraints on Galactic PeVatrons \re{remain} scarce. Recent measurements of the proton and helium spectra by LHAASO provide evidence for a distinct \emph{high-energy component}, \re{suggesting that} the knee \re{may mark the} termination of a CR population rather than simply a feature in the all-particle spectrum. At the same time, observations of ultra-high-energy (UHE) gamma rays from Galactic sources, including microquasars, provide \re{growing} evidence for particle acceleration to energies in the PeV range. In this review, we summarize the emerging picture of the CR spectrum around the knee, examine candidate Galactic PeVatrons, and discuss their possible connection to the \emph{high-energy component}. We place particular emphasis on microquasars, \re{whose demonstrated extreme particle-acceleration capabilities} and \re{potentially} sufficient \re{CR energy budget make them promising candidates for contributing to Galactic CRs around the knee}. Future measurements of intermediate-mass and heavy nuclei will be crucial for testing this picture and determining whether additional Galactic CR populations are required above the knee.}
   %{Mandatory. The objectives of the paper are defined here.} 
  % {Mandatory. The methods of the investigation are outlined here}
  % {Mandatory. The results are summarized here.}
   %{Optional, leave empty if necessary.  “Conclusions” can be used to
%explicit the general conclusions that can be drawn from the paper.}


   \maketitle

\nolinenumbers

%%%%%%%%%%%%%%%%%%%%%%%%%%%%%%%%%%%%%%%%%%%%%%%%%%%%%%%%%%%%%%
\section{Introduction: The knee after seven decades}
%\lipsum[1]
%%%%%%%%%%%%%%%%%%%%%%%%%%%%%%%%%%%%%%%%%%%%%%%%%%%%%%%%%%%%%%

\subsection{The long-standing mystery of the knee}
The ``knee'' of the Galactic cosmic-ray (CR) spectrum is one of the oldest puzzles in astroparticle physics. Since its discovery nearly seven decades ago~\citep{Kulikov1958}, extensive air shower (EAS) experiments have established a pronounced steepening of the all-particle CR spectrum at a few PeV~\citep{Nagano:1984db, Fowler:2000si, KASCADE:2005ynk, TIBETIII:2008qon, Budnev:2020oad, Ivanova:2022akw, IceCube:2019hmk}. 
Only a few years after the discovery of the knee, \citet{Peters:1961mxb} pointed out that if the break is due to the maximum energy of CR nuclei accelerated in \re{astrophysical} sources or \re{by} their propagation through interstellar space, the break energy \re{should depend on} rigidity. In other words, the \re{energies} of spectral breaks should scale approximately with the nuclear charge $Z$. Different nuclear species would therefore exhibit knee-like \re{features} at progressively higher energies, with heavier nuclei extending beyond the proton knee. This ``Peters cycle'' has provided a basic framework for interpreting the CR spectrum around and above the knee for more than six decades. 
Considerable efforts have also been devoted to measuring the CR composition around the knee.
However, identifying individual nuclear species \re{from} EAS measurements has been thought impossible until very recently. Large shower-to-shower fluctuations and uncertainties in measuring shower properties that might reflect the primary species essentially wash out the differences between showers induced by different nuclei. In addition, the hadronic interaction models used to predict EAS properties in simulations are also highly uncertain. The composition is therefore commonly characterized by the mean logarithmic mass, $\langle \ln A \rangle = \sum_i  f_i \ln A_i$, where $f_i$ is the relative fraction of nuclei with mass number $A_i$. \re{Previous measurements have indicated an increase in $\langle \ln A \rangle$ above the knee, suggesting a trend toward heavier CR composition, although with substantial uncertainties~\citep{KASCADE:2005ynk, 2012NIMPA.692...98B, IceCube:2019hmk}.} 

The physical origin of the knee, however, remains uncertain~\citep{Kachelriess:2019oqu}.
Several classes of explanations have been proposed. The knee may originate from the source population, reflecting either the maximum rigidity attainable by Galactic accelerators or an intrinsic break in the source spectrum~\citep{Stanev:1993tx, Hillas:2005cs}.
\re{The ``poly-gonato'' model by \citet{Hoerandel:2002yg, Hoerandel:2004gv} provides a phenomenological description of the knee through rigidity-dependent cutoffs in the spectra of individual nuclear species.}
Alternatively, the knee may arise from a change in the energy dependence of CR transport and escape from the Galaxy~\citep{Ptuskin1993, Candia:2002we, Candia:2003dk, Giacinti:2015hva, EspinosaCastro:2026xbs}. It has also been proposed that the CR flux around the knee is dominated by a single young nearby source, which could produce a relatively sharp spectral feature despite the expected diversity of Galactic accelerators~\citep{Erlykin:1997bs, Erlykin:2000jm, Erlykin:2003aq}. A fundamental difficulty in distinguishing these scenarios is that the knee has historically been characterized primarily through the all-particle spectrum, while the spectra of individual nuclear components were poorly constrained. Different combinations of source spectra, composition, maximum rigidity, and CR propagation could therefore produce similar features in the all-particle spectrum.

\subsection{The SNR paradigm and the search for Galactic PeVatrons}
Supernova remnants (SNRs) have long been considered the leading candidates for the origin of Galactic CRs~\citep{Baade:1934onh, Stanev:1993tx, Hillas:2005cs}. Diffusive shock acceleration (DSA) provides a natural mechanism for producing power-law CR spectra~\citep{1977DoSSR.234.1306K, 1977ICRC...11..132A, Bell:1978zc, Blandford:1978ky}, while the Galactic supernova rate and kinetic-energy budget are broadly consistent with the power required to sustain the observed CR population, assuming a CR acceleration efficiency of a few to ten percent~\citep{2010ApJ...722L..58S, Blasi:2013rva}.
Within this paradigm, SNRs are commonly expected to accelerate  Galactic CRs up to the knee~\citep{2008ARA&A..46...89R,Blasi:2013rva}.

Reaching PeV energies, however, remains theoretically challenging. 
In DSA models with magnetic-field amplification driven by the resonant streaming instability, typical estimates of the maximum proton energy are $\sim 10 - 100$~TeV~\citep{1983A&A...118..223L, 1983A&A...125..249L}. 
The discovery of the faster-growing non-resonant streaming instability~\citep{2004MNRAS.353..550B} substantially improved the prospects for PeV acceleration, particularly during the early evolution of core-collapse SNRs expanding \re{into} dense stellar winds.

Observationally, gamma-ray studies have provided evidence for hadronic CR acceleration in several SNRs, most notably through the characteristic pion-decay signature at GeV energies~\citep{FermiLAT:2009kcy, Fermi-LAT:2013iui}. Extending this evidence to PeV energies has proved substantially more difficult, and a Galactic population of SNR PeVatrons has not yet been established. Thus, although the SNR paradigm remains a well-motivated framework for the origin of Galactic CRs, whether \re{SNRs} can account for the CR population extending to the knee remains an open question.

\subsection{A new perspective on the knee problem}

The situation has changed substantially in recent years. 
\re{High-precision measurements by LHAASO place the knee at approximately 3.7~PeV, where the spectral index changes from about $2.7$ to $3.1$~\citep{LHAASO:2024knt}. In particular, LHAASO's measurements of $\langle {\rm ln A} \rangle$ suggest that the knee is primarily associated with light nuclei.}
Measurements of the proton and helium spectra around the knee provide evidence for a distinct \emph{high-energy component}~\citep{LHAASO:2025byy, LHAASO:2025mlf}. The approximate rigidity scaling of the proton and helium knee energies further suggests that these features may have a common physical origin~\citep{LHAASO:2025mlf}.
These measurements allow the knee to be \re{interpreted} not simply as a break in the all-particle spectrum, but in terms of distinct Galactic CR components~\citep{Aharonian:2026tzf}.

At the same time, ultra-high-energy (UHE) gamma-ray observations have revealed \re{numerous} Galactic sources emitting photons \re{above} $100$~TeV~\citep{LHAASO:2021gok, LHAASO:2023rpg}. The question has therefore expanded from the search for Galactic PeVatrons to identifying which \re{source} populations, if any, \re{contribute to} the \emph{high-energy component}~\citep{Aharonian:2026tzf}. In this context, microquasars have recently emerged as \re{promising} candidates, with UHE gamma-ray emission detected from several systems~\citep{LHAASO:2024psv, LHAASO:2025ysm}. Their possible connection to \re{Galactic CRs around the knee} will be a particular focus of this review.  

In the following, we first review the \re{evidence for} the \emph{high-energy component} and its physical implications. We then examine candidate Galactic PeVatron populations, with particular emphasis on microquasars. Finally, we discuss \re{future} measurements of individual CR nuclei \re{at} higher energies and \re{their} implications for the CR spectrum beyond the knee.

\section{From a spectral knee to distinct Galactic CR components}

\begin{figure*}
\centering
\includegraphics[width=\linewidth]{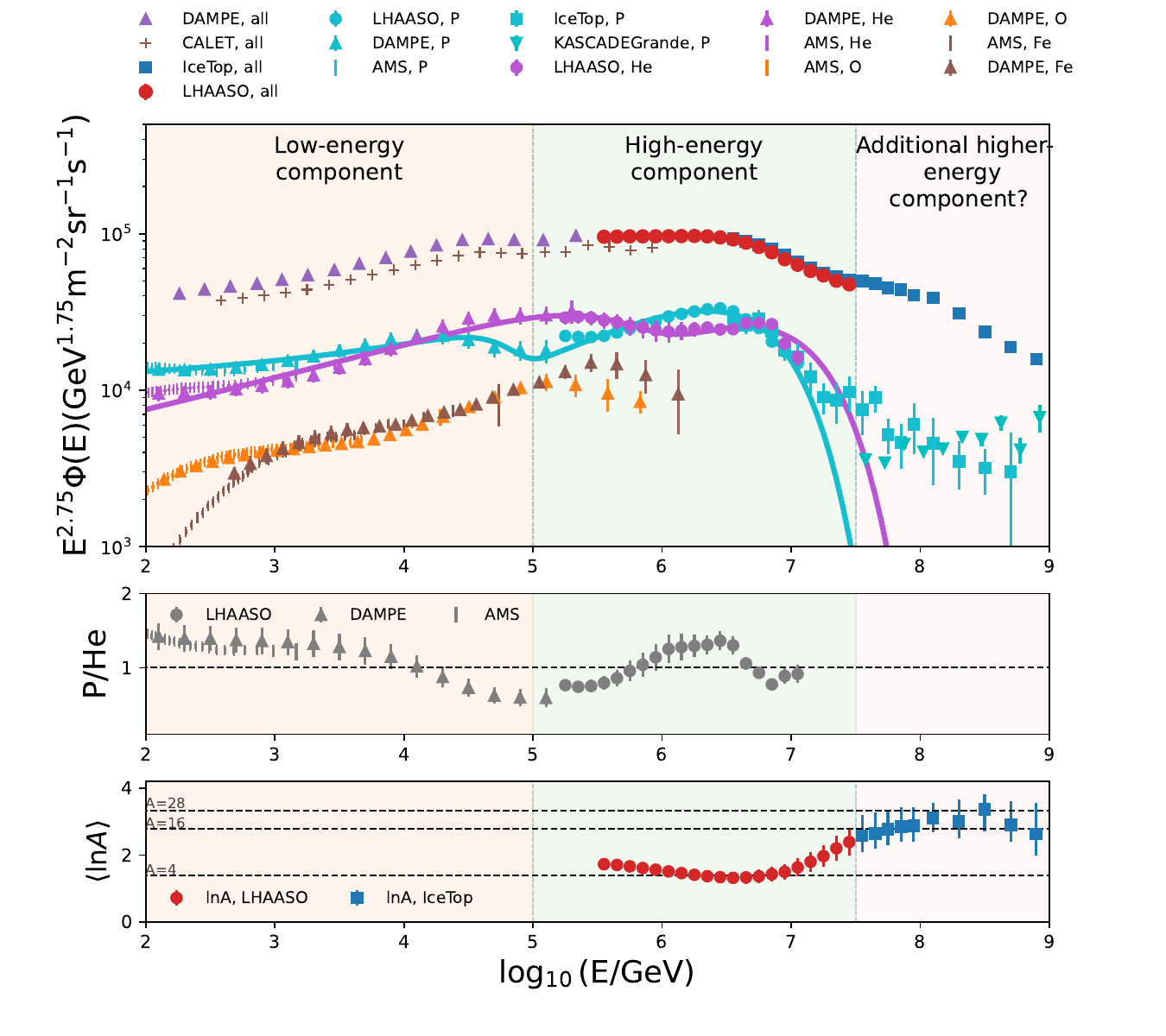}
\caption{Galactic CR spectra (top), proton-to-helium flux ratio (middle), and mean logarithmic mass, $\langle \ln A\rangle$ (bottom). The shaded regions schematically indicate phenomenological low- and high-energy components, while the hatched region indicates a possible additional component at higher energies. Their boundaries are illustrative. The all-particle spectrum is taken from~\citet{Akaike:2025UK,Cagnoli:2025mvf,LHAASO:2024knt,IceCube:2019hmk}; proton data from~\citet{AMS:2015tnn,DAMPE:2025opn,LHAASO:2025byy,IceCube:2019hmk, KASCADEGrande:2017gtn}; helium data from~\citet{AMS:2018qxq,DAMPE:2025opn,LHAASO:2025mlf}; and oxygen and iron data from~\citet{DAMPE:2025opn}. The $p/{\rm He}$ ratio is taken from~\citet{LHAASO:2025mlf}, and the $\langle \ln A\rangle$ measurements from~\citet{LHAASO:2024knt,IceCube:2019hmk}.
The solid curves for the proton and helium spectra show the two-component model of~\citet{Aharonian:2026tzf}.
}
\label{fig:cr-component}
\end{figure*}

\subsection{Spectra of individual CR nuclei around the knee}
Space-borne detectors, such as the Alpha Magnetic Spectrometer (AMS-02)~\citep{AMS:2013fma}, the DArk Matter Particle Explorer (DAMPE)~\citep{DAMPE:2017cev}, and the CALorimetric Electron Telescope (CALET)~\citep{Marrocchesi:2012nk}, \re{can} directly identify individual CR nuclei, but limited statistics at the highest energies restrict their \re{measurements} to energies below the knee. 
The hardening of the proton spectrum above several hundred GeV~\citep{AMS:2015tnn} is followed by a softening around tens of TeV, as observed by DAMPE~\citep{DAMPE:2019gys} and CALET~\citep{CALET:2022vro}.
These measurements demonstrate that the proton spectrum exhibits significant spectral structures well below the knee.
Measurements around and above the knee therefore rely primarily on EAS, for which the primary mass must be inferred indirectly from shower observables. The \re{reconstructed mass-group} spectra are \re{subject to limited} mass resolution, shower-to-shower fluctuations, and uncertainties in hadronic interaction models.

Considerable efforts have nevertheless been made to reconstruct the spectra of different nuclear components around the knee. \re{\citet{KASCADE:2005ynk} reported knee-like features in the all-particle spectrum and the spectra of light nuclei. However, reconstructing the energy spectra of five primary mass groups relied on unfolding air-shower observables with limited mass resolution and model-dependent shower simulations. In particular, the hadronic interaction models used in the analysis could not consistently describe the data over the full measured energy range, contributing to substantial uncertainties in the inferred mass-group spectra.}
%KASCADE derived energy spectra for several mass groups through an unfolding analysis of EAS observables~\citep{KASCADE:2005ynk}. 
Subsequent measurements by KASCADE-Grande~\citep{KASCADEGrande:2013lrw}, IceTop~\citep{IceCube:2019hmk}, and other EAS experiments, such as GRAPES-3~\citep{GRAPES-3:2024mhy}, further extended composition-sensitive  measurements above and below the knee.

A major advance has recently been achieved by LHAASO through the combination of electromagnetic and muon measurements from KM2A and Cherenkov measurements from WFCTA, providing complementary observables for energy reconstruction and mass discrimination. \re{This} hybrid approach enables measurements of the proton and helium spectra across the knee region, as shown in Figure~\ref{fig:cr-component}~\citep{LHAASO:2025byy, LHAASO:2025mlf}.
The helium spectrum is derived by subtracting the proton spectrum from the light-component spectrum on a consistent energy scale.
These measurements remain \re{subject to} systematic uncertainties associated with hadronic interaction models and the assumed composition of the residual background.

The proton spectrum \re{exhibits a hardening above} $\sim$0.1~PeV \re{relative to the} extrapolation from lower \re{energies, followed by a steepening} at $\sim3.3$~PeV, close to the all-particle knee at $\sim 3.7$~PeV.
Its spectral index changes from approximately 2.5 below the knee to 3.5 above it~\citep{LHAASO:2025byy}.
The helium spectrum shows a hardening around 1.1~PeV, followed by a softening around 7~PeV~\citep{LHAASO:2025mlf}. 
The proton and helium knee energies are compatible with rigidity scaling within uncertainties, although the lower-energy hardening features do not exhibit the same simple scaling.
The $p/{\rm He}$ flux ratio at the same total particle energy crosses unity near $0.7$~PeV as protons overtake helium, and again near 5~PeV as helium becomes more abundant than protons~\citep{LHAASO:2025mlf}. 
These changes reveal substantial structure in the individual nuclear spectra that is not apparent from the relatively smooth all-particle spectrum alone.

\subsection{Emergence of the high-energy component}

The LHAASO proton spectrum suggests an additional contribution toward PeV energies, which we refer to phenomenologically as the \emph{high-energy component}~\citep{LHAASO:2025byy}. 
When the helium spectrum is considered together with the proton spectrum, \re{their} knee energies are consistent with approximate rigidity scaling within the reported uncertainties~\citep{LHAASO:2025mlf}, suggesting that the particles near the two knees may originate from a common CR population. 
Furthermore, the evolution of the proton-to-helium ratio, $p/{\rm He}$, as shown in the middle panel of Figure~\ref{fig:cr-component}, suggests that the composition of the \emph{high-energy component} may differ from that of the low-energy component. 

\re{Multiple source populations have also been proposed to explain the spectra of individual CR nuclei and the all-particle spectrum~\citep[e.g.,][]{Zatsepin:2006ci, Gaisser:2013bla}. Some scenarios invoke one or more nearby sources to explain spectral features around tens of TeV~\citep[e.g.,][]{Lv:2024wrs, Qiao:2025zyk, Yuan:2025xqt, Li:2026uox}. In such models, individual source populations may follow rigidity-dependent cutoffs, while their superposition produces more complex proton and helium spectra.} 
\re{Nevertheless, the current proton spectrum can also be described without invoking a dominant contribution from a nearby source, although such a contribution cannot be excluded.}
\re{Motivated by a minimal description of the data}, \citet{Aharonian:2026tzf} proposed a minimal two-component interpretation of the Galactic CR proton and helium spectra from GeV to PeV energies, as illustrated in Figure~\ref{fig:cr-component}. In this picture, the \emph{high-energy component} can be described by rigidity-dependent proton and helium spectra with a common spectral index, $\Gamma \sim 2.4$, and a characteristic cutoff rigidity corresponding to a proton cutoff energy of $\sim 6.2$~PeV.  
In contrast, the low-energy proton and helium components do not follow simple rigidity scaling. In this fit, the low-energy proton component cuts off at $\sim 70$~TeV with a relatively sharp cutoff, whereas the helium component extends to $\sim 450$~TeV with a smoother, approximately exponential cutoff. \re{ Their different cutoff rigidities may reflect contributions from different SNR populations or different acceleration sites or environments within SNRs, as discussed by \citet{Aharonian:2026tzf}.}
Within this two-component interpretation, the spectral structure observed from tens to hundreds of TeV can be understood as the transition between the low- and high-energy components, while the knee at PeV energies is primarily associated with the termination of the \emph{high-energy component}. 
The model also accounts for the decrease and subsequent recovery of the $p/{\rm He}$ ratio through the changing relative contributions of the two components.
This decomposition provides a phenomenological description of the measured spectral structures rather than a direct identification of their astrophysical origins.

The evolution of the mean logarithmic mass shown in the lower panel of Figure~\ref{fig:cr-component} provides complementary information on the changing composition.
LHAASO measurements show that $\langle \ln A \rangle$ decreases from approximately $1.7$ at $0.3$~PeV to $1.3$ near $3$~PeV and then increases above the knee~\citep{LHAASO:2024knt}. This evolution indicates an increasing relative contribution from light nuclei below the knee, followed by a shift toward heavier nuclei as the light-component spectra steepen. 
At higher energies, IceCube/IceTop measurements indicate a progressively heavier composition up to approximately $100$~PeV~\citep{IceCube:2019hmk}. These measurements constrain the overall composition trend, although the inferred $\langle \ln A \rangle$ remains sensitive to hadronic interaction models. Moreover, this quantity alone does not uniquely determine the individual nuclear spectra.

The approximate rigidity scaling of the proton and helium knee energies provides a natural starting point for interpreting this evolution. If the \emph{high-energy component} extends to heavier nuclei with a characteristic maximum energy proportional to charge,
$E_{\max,Z}\simeq Z E_{\max,p}$,
heavier nuclei would extend progressively beyond the proton and helium knees. 
Measurements of intermediate-mass nuclei and, especially, the iron spectrum would test whether the trend toward a heavier composition reflects this scaling.

\subsection{\re{Complementary constraints from gamma rays and CR anisotropy}}
\re{The CR spectrum measured in the solar neighborhood does not necessarily represent the CR distribution throughout the Galaxy. Galactic diffuse gamma-ray emission provides an important probe of CR populations beyond the solar neighborhood. LHAASO has measured diffuse VHE--UHE gamma-ray emission from both the inner and outer Galactic plane~\citep{LHAASO:2023gne, LHAASO:2024lnz}. The measured fluxes exceed predictions based on locally measured CR spectra and Galactic gas distributions, suggesting possible spatial variations in the CR density. The more pronounced excess in the inner Galaxy may also indicate additional contributions from unresolved sources. 
Most recently, LHAASO detected VHE gamma-ray emission from five nearby giant molecular clouds through a stacking analysis, providing an independent probe of CR spectra away from Earth. Although no significant CR knee signature was identified, the measurements are broadly consistent with a proton knee at PeV energies~\citep{LHAASO:2025smo}. The detection of Galactic diffuse neutrinos supports the presence of hadronic CR interactions in the Galaxy~\citep{IceCube:2023ame, IceCube:2026plr}. However, current measurements provide limited constraints on Galactic neutrino emission at the higher energies. Future observations with upgraded or next-generation neutrino detectors will be important for probing Galactic PeVatrons.
}

\re{Measurements of CR anisotropy provide complementary constraints on the spatial distribution of CR sources and their propagation through Galactic magnetic fields. Recent LHAASO observations have revealed energy-dependent shifts in medium-scale CR anisotropies above 10~TeV, highlighting the role of local magnetic turbulence in shaping the observed arrival directions~\citep{LHAASO:2025dsq}. Models in which a small number of sources dominate the local PeV flux should therefore be tested against both the CR spectrum and the amplitude and phase of the large-scale anisotropy.}

\section{Galactic PeVatrons and the origin of the high-energy CR component}

Evidence for a \emph{high-energy component} naturally raises the question of its astrophysical origin. This question has become particularly timely with the emergence of UHE gamma-ray astronomy~\citep{LHAASO:2021gok, Cao:2023mig}. LHAASO has revealed a population of Galactic sources emitting gamma rays beyond $100$~TeV, associated with different classes of astrophysical objects~\citep{LHAASO:2023rpg}.

\re{Identifying the sources of the Galactic CR knee requires connecting their acceleration capabilities to the observed CR spectra and composition. Table~\ref{tab:pevatron-knee} summarizes
the current status of four major
candidate source classes.}

\begin{table*}
\centering
\caption{\re{Comparison of Galactic source
candidates for the CR knee. See the text for references and discussion.}}
\label{tab:pevatron-knee}

\footnotesize
\setlength{\tabcolsep}{5pt}
\renewcommand{\arraystretch}{1.30}
\setlength{\aboverulesep}{0pt}
\setlength{\belowrulesep}{0pt}

\begin{tabularx}{\textwidth}{
>{\raggedright\arraybackslash}p{2.0cm}
YYYY
}
\toprule
\rowcolor{tableheader}
\textbf{Source class} &
\textbf{UHE $\gamma$ rays} &
\textbf{PeV protons} &
\textbf{Population / power} &
\textbf{$p$ and He knee modeling}\\
\midrule

\rowcolor{tablealt}
\textbf{Microquasars} &
Several detections;
Cyg X-3: $\sim 4$~PeV &
Supported in
hadronic models &
Potentially sufficient;
population uncertain &
Proton and helium
spectra modeled
\\

\textbf{SNRs} &
UHE emission near SNRs;
origin debated &
Possible in
favorable conditions &
Numerous;
PeV subset uncertain &
Extensively modeled; PeV knee uncertain
\\

\rowcolor{tablealt}
\textbf{Pulsars/PWNe} &
Numerous;
including PeV &
Hadronic acceleration
uncertain &
Numerous;
hadronic power uncertain &
No established joint hadronic fit
\\

\textbf{YMSCs} &
Cygnus Bubble;
association uncertain &
Possible in
favorable conditions &
Population uncertain; 
PeV efficiency uncertain &
Joint knee modeled; contribution uncertain
\\

\bottomrule
\end{tabularx}
\end{table*}

\subsection{Microquasars (or X-ray binary jets)}

Microquasars are X-ray binary systems in which a compact object, either a black hole (BH) or a neutron star (NS), accretes matter from a companion star and launches relativistic jets~\citep{Mirabel:1999fy, 2014SSRv..183..323F, Olivera-Nieto:2026jxb}.

The possible role of microquasars as Galactic CR accelerators has a long history. Early studies considered relativistic outflows from accreting compact objects as potential sites of high-energy particle acceleration~\citep{1984Natur.312...50H, Heinz:2002qj, Romero:2003td, Bednarek:2005gf, Cooper:2020tzq, Escobar:2022jod, Kantzas:2023oww}.

\re{Subsequent studies investigated the potential contribution of X-ray binary jets to Galactic CRs.} \citet{Cooper:2020tzq} \re{proposed that X-ray binary jets could provide an additional Galactic CR component, potentially dominating the spectrum above the knee, particularly at} $\sim10$--$100$~PeV.
\citet{Kantzas:2023oww} \re{investigated the contribution of} a population of $\sim$1000 black-hole X-ray binaries, adopting a CR injection spectral index of about $2.2$. In their \re{model}, low-mass X-ray binaries were assumed to accelerate protons up to $\sim$100 TeV and high-mass X-ray binaries up to $\sim1$ PeV, while maximum energies as high as $\sim10$~PeV were also considered.
These studies \re{suggested} that X-ray binary jets could contribute to Galactic CRs, \re{particularly at high energies}. However, their predictions depend strongly on assumptions about the acceleration spectrum, maximum particle energy, source population, and CR normalization. \re{They did not specifically address the proton and helium spectral structures now measured around the knee by LHAASO.}

The situation has changed with the
advent of UHE gamma-ray observations.
LHAASO has detected UHE gamma-ray
emission from several microquasar
systems~\citep{LHAASO:2024psv,
LHAASO:2025ysm,LHAASO:2026gzg}.
\re{Some of these sources exhibit
broadly comparable photon indices,
with emission extending to several
hundred TeV and, in some cases,
into the PeV range.
These observations provide strong
evidence for extreme particle
acceleration in microquasars,
although the underlying hadronic CR
spectra and escape properties
remain uncertain.}

Together with the evidence for the \emph{high-energy component} discussed in the previous section, these observations motivate \re{the possibility that} microquasars constitute an important source population \re{contributing to the \emph{high-energy component}} and \re{shaping} the knee structure in the all-particle spectrum~\citep[e.g.,][]{LHAASO:2024psv, Yue:2024nns, Ohira:2024qtr, Zhang:2025tew, Wang:2025yqy, Kaci:2025gyb, Bykov:2025xxl, Zhang:2026igt, Aharonian:2026tzf, Nie:2026ykh, Vecchiotti:2026okk, Yu:2026vco}.

As discussed in the previous section, the \emph{high-energy component} has a characteristic proton cutoff energy of approximately $6$~PeV. If this feature reflects the maximum rigidity of the underlying source population, the relevant Galactic accelerators must be capable of accelerating protons to at least several PeV.
For a Galactic population of a few to tens of active microquasars with typical jet powers of $L_j\sim10^{38}$--$10^{39}\ {\rm erg\,s^{-1}}$, \re{a CR acceleration efficiency of order $10\%$ could, in principle, provide a CR power comparable to that required for PeV CRs~\citep{LHAASO:2024psv, Zhang:2025tew}. This estimate depends on the source duty cycle and the fraction of accelerated-particle power released into this energy range}.
\re{\citet{Kaci:2025gyb} showed that a population of approximately ten active, powerful microquasars could account for the observed proton spectrum around the knee and the Galactic diffuse gamma-ray emission under their adopted assumptions.
\citet{Vecchiotti:2026okk} investigated the contribution of observed microquasars, using gamma-ray measurements to constrain their hadronic CR luminosities. They estimated that the detected sources could contribute up to $\sim4\%$ of the local CR flux at 1~PeV \re{under} their benchmark assumptions, increasing to $\sim37\%$ for higher acceleration efficiencies.
These studies highlight the importance of the source population, CR injection history, and propagation in determining the contribution of microquasars \re{to Galactic CRs}.}

A key question is whether microquasars can accelerate hadronic CRs to the required PeV energies.
Several acceleration mechanisms have been proposed, including shocks and shear acceleration in large-scale jets~\citep{Zhang:2025tew,Wang:2025yqy}. In the shear-acceleration scenario of \citet{Zhang:2025tew}, pre-existing Galactic CRs can be reaccelerated in microquasar jet--cocoon systems, producing a hard escaping spectrum extending to PeV energies. For characteristic jet and cocoon parameters, the maximum proton energy can reach several PeV, \re{as} required for the \emph{high-energy component}.
\re{In this scenario, the spectra of protons, helium, and heavier nuclei are predicted from the reacceleration of pre-existing Galactic CRs, with their relative abundances linked to the composition at the injection rigidity~\citep{Zhang:2025tew}. Future nuclei-resolved measurements will provide further tests of these predictions.}

\re{Observationally}, Cygnus X-3 provides evidence for extreme particle acceleration, with gamma-ray emission extending into the PeV regime~\citep{LHAASO:2025ysm}.
\re{Its} spectral hardening toward PeV energies can be interpreted in terms of photomeson production, \re{supporting a hadronic origin}~\citep{LHAASO:2025ysm, Wei:2025zkm, Kachelriess:2025jtm, Zdziarski:2026jbu, Wan:2026xtp}.
GRS 1915+105 provides another example, for which a hadronic interpretation of the observed GeV--PeV gamma-ray emission is favored~\citep{LHAASO:2026gzg}.
These observations \re{provide growing support for} hadronic PeV \re{acceleration in microquasars}. Together with their potentially sufficient energy budget, these make microquasars \re{promising} candidates for the \emph{high-energy component}.

\re{Another} important aspect is the propagation of CRs injected by microquasars. The observed CR spectrum reflects the cumulative contribution of past source activity over the energy-dependent CR residence time. The spatial and temporal distribution of past microquasar activity, together with transport in the large-scale ordered and turbulent Galactic magnetic fields, must therefore be considered.
\citet{Zhang:2026igt} investigated the contribution of microquasar remnants, \re{namely} the CR populations left by past microquasar activity, to the proton spectrum around the knee, accounting for their propagation through the Galactic magnetic field. They found that a bump-like feature around the knee can arise for sufficiently hard CR injection spectra, with $s_{\rm cr} \lesssim 2$~\citep[see also][]{Zhang:2025tew, Kaci:2025gyb}.
\re{In their simulations, the local PeV CR flux is often dominated by a small number of remnants, with as few as four sources accounting for approximately half of the observed flux at 1~PeV. Individual source contributions depend strongly on their magnetic connectivity to the solar neighborhood, rather than simply on their distance.}
\re{These results suggest that microquasars may contribute significantly to the CR flux around the knee, although uncertainties in the source population, CR injection and escape, and Galactic propagation remain.}

\subsection{\re{Other Galactic PeVatron candidates}}

Other classes of Galactic PeVatrons may also contribute to the inferred \emph{high-energy component}. As with microquasars, establishing a direct connection between these populations and the CR knee requires further observational evidence. 

\paragraph{Supernova remnants} SNRs have favorable energetics and well-established evidence for hadronic acceleration, but their ability to accelerate a substantial CR population to PeV energies remains uncertain. 
\re{Recent gamma-ray observations of young SNRs, such as Cas A, have not revealed UHE gamma-ray emission, placing important constraints on ongoing PeV particle acceleration~\citep{LHAASO:2023igp, LHAASO:2025xmn}.
However, even if PeV energies are reached during the early evolution phase, only a small fraction of the total kinetic energy may \re{be transferred to} such high-energy particles. As a result, the time-integrated CR output at PeV energies \re{may} remain small~\citep{Blasi:2025kcy}, \re{making} it challenging for a typical SNR population to account for the CR flux around the knee.}
\re{Gamma-ray observations of middle-aged SNRs, including W51C~\citep{2024SciBu..69.2833C}, IC 443~\citep{LHAASO:2025aoo}, G150.3+4.5~\citep{LHAASO:2026irn}, and $\gamma$-Cygni~\citep{LHAASO:2026irn}, provide important constraints on CR acceleration and escape. These observations may not directly constrain the maximum particle energies reached during the early SNR evolution, \re{as} the highest-energy CRs may have already escaped. A Galactic population of hadronic SNR PeVatrons capable of accounting for the CR knee has yet to be established observationally.}

\re{Several scenarios may nevertheless allow a subset of SNRs or supernovae to accelerate CRs to PeV energies. \citet{Cristofari:2020mdf} already showed that rare, energetic core-collapse supernovae, particularly their Type II$^*$ scenario with explosion energies of order $10^{52}\rm~erg$, could accelerate protons to PeV energies and contribute to the observed CR spectrum. More recently, \citet{Bell:2025jbi} proposed that magnetic mirroring could enhance CR confinement near SNR shocks, potentially allowing acceleration to PeV energies. Interacting supernovae in  dense circumstellar environments, particularly Type IIn events, have also been proposed as important contributors to CRs from a few PeV to hundreds of PeV~\citep{Ekanger:2026coc}. 
}
\re{From a population perspective, \citet{Evoli:2026rpj} showed that variations in the maximum rigidity among SNRs can produce a knee-like feature. Their fit to the LHAASO proton spectrum requires a relatively narrow distribution of maximum energies, favoring a restricted subset of SNRs rather than the full Galactic SNR population. These studies highlight possible pathways to PeV acceleration within the SNR framework, although their contributions to Galactic CRs remain uncertain and require further observational constraints.}
\re{
Although proton and helium spectra
have been extensively studied within
the SNR framework, reproducing the
detailed spectral features now measured
by LHAASO remains an important test
of these models.
}

\paragraph{Pulsars and pulsar wind nebulae} 
Pulsars and pulsar wind nebulae (PWNe) are firmly established as extreme particle accelerators~\citep{LHAASO:2021cbz, LHAASO:2026gno}, and many UHE gamma-ray sources are associated with these systems~\citep{LHAASO:2023rpg}. However, their UHE gamma-ray emission can often be explained by leptonic processes, while evidence for a substantial population of escaping hadronic CRs remains limited~\citep{Lu:2026pey}. Their connection to the \emph{high-energy component} is therefore less direct.  

\paragraph{Young massive stellar clusters and superbubbles}
Young massive stellar clusters (YMSCs) and superbubbles provide another possible environment for accelerating CRs to PeV energies, with the Cygnus Bubble \re{as} a prominent example \re{of an extreme CR acceleration region} ~\citep{LHAASO:2023uhj}. However, accelerating CRs beyond PeV energies through stellar winds alone remains theoretically challenging, and more complex scenarios, such as supernova explosions occurring within stellar clusters, may also contribute~\citep{Haerer:2025ull, Aharonian:2026tzf, Qiu:2026kdu}.
The Galactic YMSC population and,
in particular, the fraction capable
of accelerating and releasing PeV
CRs remain uncertain.

Other source classes, such as isolated black holes embedded in dense environments, have also been proposed as Galactic CR accelerators~\citep{Kimura:2024izc}. Their contribution to the \emph{high-energy component}, however, remains highly uncertain.

\section{Discussion}

\subsection{\re{Implications of precise CR measurements for source models}}

\re{The precise proton and helium spectra measured by LHAASO provide quantitative constraints on models of the Galactic CR knee.
The proton spectrum steepens at $\sim 3.3\rm~PeV$, with the spectral index changing from approximately 2.5 to 3.5, while the helium knee occurs at $\sim 7\rm~PeV$.
A viable source model must reproduce these spectral features and their approximate rigidity scaling, rather than merely demonstrate the ability to accelerate particles to PeV energies.
In acceleration-limited scenarios, the relatively sharp proton knee also constrains the distribution of maximum rigidities among contributing sources.  
However, the observed CR spectra depend
not only on particle acceleration,
but also on CR escape and Galactic
propagation.
Connecting the observed Galactic
PeVatrons to the CR knee therefore
requires a consistent treatment of
these processes.
}

\re{More generally, the approximately power-law Galactic CR spectrum can arise from the collective
contribution of numerous
sources with different locations,
injection spectra, and maximum energies.
Such a superposition can produce a relatively smooth spectrum without requiring identical source properties. 
Nevertheless, individual nearby sources may
still contribute significantly at certain energies and produce local spectral features. 
The contribution of individual sources may become more prominent at higher energies, particularly around the knee, where
fewer sources may reach the required energies. The observed spectrum may therefore become more sensitive to
 individual source histories, their spatial distribution, 
and propagation through Galactic
magnetic fields.}

\re{
The detection of gamma rays up to
$\sim 4$~PeV from Cygnus X-3
suggests proton acceleration to
several tens of PeV in hadronic
scenarios~\citep{LHAASO:2025ysm}.
If heavy nuclei are present, can be accelerated to comparable maximum rigidities, and survive
interactions within the source,
iron nuclei could potentially reach
energies approaching $1$~EeV.
Such extreme Galactic
accelerators may therefore provide valuable clues to the upper end of Galactic CR acceleration and the transition
toward extragalactic CRs.
}

\subsection{The second knee and \re{possible additional CR} components}

%The identification of a distinct \emph{high-energy component} also provides a new perspective on the CR spectrum beyond the knee. 
The all-particle spectrum exhibits additional structures between the knee and the ankle, most notably the second knee at around 100~PeV~\citep[e.g.,][]{HiRes-MIA:2000ook}. 
\re{Its origin} remains uncertain~\citep{Kachelriess:2019oqu}. 

If the proton knee at $\sim 3.3\rm~PeV$ reflects a rigidity-dependent spectral feature of the \emph{high-energy component}, a corresponding steepening in the iron spectrum would be expected at  approximately $26\times 3.3 \simeq 86$~PeV.
Note that the observed proton knee energy
differs from the exponential cutoff
parameter of $\sim 6.2$~PeV adopted in
the phenomenological model of
\citet{Aharonian:2026tzf}, since the
two values characterize the spectral
steepening using different functional forms.
However, this iron knee should not be directly
identified with the second knee in
the all-particle spectrum.
The latter may reflect the superposition or transition of different CR populations.
Measurements of intermediate-mass
and heavy nuclei are essential for
distinguishing these possibilities.

In addition to the second knee,
several measurements \re{have} indicated a hardening of the proton or light component \re{at energies around $\sim100$ PeV}~\citep{Apel:2013ura, IceCube:2019hmk}. 
\re{This feature may point to an additional CR population with} a substantially higher maximum rigidity~\citep{Zhang:2025tew}. 
In the phenomenological interpretation of \citet{Aharonian:2026tzf}, an additional proton component may extend to energies approaching or exceeding 100~PeV, \re{well beyond} the cutoff of the previously identified \emph{high-energy component}. 
\re{Its origin remains uncertain and could be Galactic or extragalactic~\citep{Zhang:2025tew}.}

If Galactic, \re{such a component would require accelerators} capable of reaching \re{substantially} higher rigidities \re{than those responsible for the conventional knee}. Possible candidates include extreme microquasars, energetic stellar explosions, or accelerators associated with the supermassive black hole (SMBH) at the Galactic center. 
The nuclear composition of this
additional component may also differ
from that of the CR population
dominating around the conventional
knee.
%Cygnus X-3 provides an interesting example of an extreme compact system: its PeV gamma-ray emission \re{indicates extreme particle acceleration}, while its Wolf-Rayet companion provides a helium- and CNO-rich environment. Such systems therefore motivate the possibility that \re{an additional} higher-energy Galactic component could have a nuclear composition different from that of the CR population dominating around the conventional knee. 

\subsection{Challenges and opportunities for nuclei-resolved measurements at higher energies}

Extending measurements of individual nuclear species to higher energies will be crucial for distinguishing these scenarios. 
Measurements of intermediate-mass and iron nuclei will be especially important for \re{testing the rigidity scaling of the \emph{high-energy component} and distinguishing its possible iron knee from the second knee in the all-particle spectrum.} At the same time, improved measurements of the proton and helium components above the knee are needed to determine whether an additional Galactic source population with a higher rigidity cutoff emerges at higher energies.

One \re{possible} approach is to turn the LHAASO-WFCTA into a fluorescence telescope array to watch over the space above the LHAASO muon detector array for showers at energies above 100 PeV. Combining the LHAASO-KM2A muon detectors, such a configuration could retain sensitivity to primary composition and could improve measurements of \re{nuclear mass} groups covering the second knee. Complementary measurements from next-generation experiments, such as the IceCube-Gen2 surface array, are also expected to improve CR composition measurements in the energy range relevant to the transition from Galactic to extragalactic CRs.

\section{Summary}
The origin of the Galactic CR knee has remained a long-standing challenge, largely because \re{studies} of individual nuclear spectra around the knee \re{have long been} troubled by the very limited precision \re{of available measurements}. At the same time, the physical properties of Galactic PeVatrons were poorly constrained observationally, leaving open whether the knee is produced by \re{a subset of} extreme SNRs or by other classes of Galactic accelerators. 
The situation has changed substantially with the operation of LHAASO. Measurements of the \re{proton and helium spectra} around the knee, together with the discovery of a population of Galactic UHE gamma-ray sources, now provide \re{an} opportunity to \re{investigate} the knee from both the CR and source perspectives. 
The proton and helium spectra provide evidence \re{for an additional contribution around the knee, phenomenologically described as a} distinct \emph{high-energy component}, with knee energies consistent with rigidity scaling. In this picture, the Galactic CR knee can be interpreted as being primarily \re{associated with the termination of this component, possibly reflecting a characteristic maximal rigidity of the \re{underlying} CR population}, rather than simply as a break in the all-particle spectrum.
In parallel, recent UHE gamma-ray observations have established microquasars as an important class of Galactic PeVatron candidates. Their \re{ability to accelerate particles to PeV} energies, \re{together with their potentially} sufficient CR \re{energy budget}, makes them \re{compelling} candidates for contributing to the \emph{high-energy component} and thereby shaping the Galactic CR knee.

Looking beyond the knee, future measurements of intermediate-mass and heavy nuclei will be essential \re{to fully understanding} whether rigidity scaling \re{holds} throughout the full Peters cycle and whether \re{the emergence of} an extragalactic component \re{is manifested} in the \re{measured} spectra. 
Improved measurements of the proton and light components at higher energies will also \re{help determine} whether an additional \re{CR population with a higher maximum rigidity} is required \re{and constrain its possible Galactic or extragalactic origin}. 

%%%%%%%%%%%%%%%%%%%%%%%%%%%%%%%%%%%%%%%%%%%%%%%%%%%%%%%%%%%%%%
\begin{acknowledgements}
B.T.Z. is supported in China by National Key R\&D program of China under the grant 2024YFA1611402.
During the preparation of this work, the authors used ChatGPT (OpenAI) to assist with language editing, manuscript organization, and checks of clarity and consistency. After using this tool, the authors reviewed and edited the content as needed and take full responsibility for the content of the published article.
\end{acknowledgements}

% Please note that we have included the references below in
% order to compile the document, but we ask you to:
\bibliographystyle{PHEA} % style aa.bst
\bibliography{main} % your references Yourfile.bib

% %%%%%%%%%%%%%%%%%%%%%%%%%%%%%%%%%%%%%%%%%%%%%%%%%%%%%%%%%%%%%%
% Example below of non-structurated natbib references  
% To use the v8.3 macros with this form of composition of bibliography,
% the option "bibyear" should be added to the command line
% "\documentclass[bibyear]{aa}".
% %%%%%%%%%%%%%%%%%%%%%%%%%%%%%%%%%%%%%%%%%%%%%%%%%%%%%%%%%%%%%%

% \begin{thebibliography}{}

%   \bibitem[1966]{baker} Baker, N. 1966,
%       in Stellar Evolution,
%       ed.\ R. F. Stein,\& A. G. W. Cameron
%       (Plenum, New York) 333

%    \bibitem[1988]{balluch} Balluch, M. 1988,
%       A\&A, 200, 58

%    \bibitem[1980]{cox} Cox, J. P. 1980,
%       Theory of Stellar Pulsation
%       (Princeton University Press, Princeton) 165

%    \bibitem[1969]{cox69} Cox, A. N.,\& Stewart, J. N. 1969,
%       Academia Nauk, Scientific Information 15, 1

%    \bibitem[1980]{mizuno} Mizuno H. 1980,
%       Prog. Theor. Phys., 64, 544
   
%    \bibitem[1987]{tscharnuter} Tscharnuter W. M. 1987,
%       A\&A, 188, 55
  
%    \bibitem[1992]{terlevich} Terlevich, R. 1992, in ASP Conf. Ser. 31,
%       Relationships between Active Galactic Nuclei and Starburst Galaxies,
%       ed. A. V. Filippenko, 13

%    \bibitem[1980a]{yorke80a} Yorke, H. W. 1980a,
%       A\&A, 86, 286

%    \bibitem[1997]{zheng} Zheng, W., Davidsen, A. F., Tytler, D. \& Kriss, G. A.
%       1997, preprint
% \end{thebibliography}

\end{document}